\documentclass[runningheads,a4paper]{llncs}

\usepackage[caption=false]{subfig}
\usepackage{amssymb}
\usepackage{graphicx}
\usepackage{algorithmic}
\usepackage{algorithm}
\usepackage{float}
\usepackage{amsmath}
\newfloat{algorithm}{t}{lop}

\usepackage{url}
\newcommand{\keywords}[1]{\par\addvspace\baselineskip
\noindent\keywordname\enspace\ignorespaces#1}

\include{Qcircuit}

\begin{document}

\mainmatter  

\title{Classical Control of Large-Scale Quantum Computers}


\author{Simon J. Devitt$^{1,2}$}
\authorrunning{Classical Control of Large-Scale Quantum Computers}

\institute{$^1$Ochanomizu University, 
2-1-1 Otsuka, Bunkyo-ku, Tokyo 112-8610, Japan\\
$^2$ National Institute of Informatics, 2-1-2 Hitotsubashi, Chiyoda-ku, Tokyo, Japan
}

%
%

\toctitle{Lecture Notes in Computer Science}
\tocauthor{Authors' Instructions}
\maketitle

\begin{abstract}
The accelerated development of quantum technology has reached a pivotal point.  Early in 2014, several results were published 
demonstrating that several experimental technologies are now accurate enough to satisfy the requirements of fault-tolerant, 
error corrected quantum computation.  While there are many technological and experimental issues that still need to be solved, 
the ability of experimental systems to now have error rates low enough to satisfy the fault-tolerant threshold for several error 
correction models is a tremendous milestone.  Consequently, it is now a good time for the computer science and 
classical engineering community to examine the {\em classical} problems associated with compiling quantum algorithms 
and implementing them on future quantum hardware.  In this paper, we will review the basic operational rules of 
a topological quantum computing architecture and outline one of the most important classical problems that need to be 
solved; the decoding of error correction data for a large-scale quantum computer.  
We will endeavour to present these problems independently from the underlying physics as much of this work  
can be effectively solved by non-experts in quantum information or quantum mechanics.

\end{abstract}

\section{Introduction}
Quantum technology, specifically large-scale quantum computation, has been a significant research topic in physics since 
the early 1990's.  Since the publication of the first quantum algorithms \cite{NC00}, illustrating the computational power of 
quantum computers, millions of dollars has been invested worldwide and numerous technological advances have been made 
\cite{G06,HA08,PLZY08,PMO09,P12,L12}.  It is now routine for multiple experimental laboratories to fabricate and 
control small arrays of quantum bits (qubits) 
and perform proof of principal experiments demonstrating small quantum algorithms and protocols \cite{L10}.  Quantum technology 
has also moved into the industrial sector via protocols such as Quantum Key Distribution (QKD) and Quantum random number 
generators and many non-physicists are aware of the D-Wave quantum computer, which while scientifically controversial is an 
attempt to build a analogue quantum computer capable of solving certain types of optimisation problems \cite{VAMWL14,BRIWWLMM13,SSSV14}.  

Recent experimental results in 2014 have demonstrated that two experimental systems can be built with high enough accuracy 
to satisfy the constraints of fault-tolerant, error corrected quantum computation \cite{B14,CDMF14}.  As error rates on qubit arrays is high compared to 
classical nano-electronics, extensive error correction is required to successfully perform computation 
\cite{NC00,DMN13,S03+,CRSS98,KLV99}.  
One of the most 
seminal results in quantum information theory is the fault-tolerant threshold theorem \cite{AB97}.  This theorem states that provided 
the fundamental error rate associated with qubits and quantum gates falls below a threshold, then arbitrarily long quantum 
computation is possible with a polylogarithmic overhead in physical resources.  This threshold is a function of the type 
of quantum error correction code used for the computer \cite{DMN13,S03+,CRSS98,KLV99} 
and extensive research has been performed to derive new codes, 
with high thresholds, that are amenable to experimental architectures.  
Arguably the most successful class of codes that 
have been developed are known as topological quantum codes  \cite{K97,DKLP02,RHG06,RH07,FMMC12}.  
Topological quantum codes are defined over a lattice 
(of arbitrary dimension depending on the code, but the most common are 2- and 3-dimensional) of physical qubits.  The code 
itself can be defined over small, physically local groups of qubits while the properties of the encoded information is a global 
property of the entire lattice.  This is what defines the code as topological.  These codes are arguably preferred in quantum 
computer development as they exhibit comparably high fault-tolerant thresholds and they are adaptable to the physical 
constraints of experimental quantum systems.  

Irrespective of the actual quantum code chosen to protect a quantum computer, it is well known that operating such as 
system requires extensive {\em classical} control infrastructure.   This is not simply related to the control of the physical 
device hardware needed to operate a qubit (lasers, signal generators etc...), but it is also required to decode error correction 
information produced by the computer.  This classical control software development is 
in its infancy and has received little attention within the fields of quantum information and classical computer 
science \cite{DFTMN10,GP14}.  While 
there has been much work at the more abstract level of quantum algorithm design and circuit optimisation \cite{KMM13,VI05,Z98,VBE96,CV11,WC00,MI05}, we now have to 
go one step deeper and connect the high level work to the physical constraints of the quantum hardware.  

This paper will introduce one of the main classical computer science and engineering problems associated with controlling 
a large scale quantum computer.  We will focus on a specific form of quantum 
computer; namely a system that is built using an error correction code known as Topological Quantum Clusters (TQC) \cite{RHG07,FG08}.  This 
code has received significant attention in recent years due to multiple hardware architectures utilising it in designing 
large scale systems \cite{FMMC12,SJ08,DFSG08,NTDS13,JMFMKLY10,MF10,MRRBMD14}.  We won't discuss the 
details of how information to be encoded or manipulated.  Instead we will focus on the basic error correction properties of the code and 
what this implies for classical processing of this data.  In section \ref{sec:background} we will provide some background information on the basic 
definitions of qubits and quantum logic.  In section \ref{sec:topological} we will provide a brief introduction to the TQC model.  This 
will not be an in depth introduction, but should provide enough material to grasp the classical problems that need to be solved.  
Finally, in section \ref{sec:errorcorrection} we will examine the processing that needs to be developed to perform 
dynamic error correction on the system and discuss the potential problems associated with the massive amount of 
classical data produced by the computer. 

\section{Quantum Computers}
\label{sec:background}

A qubit is the quantum analogue of a bit. Its state is defined as a vector of
dimension $2$, where $\ket{0}=(1, 0)^T$ is the vector notation for the value corresponding
to binary $0$, and $\ket{1}=(0 , 1)^T$ correspond to $1$. The state of one qubit $q$ can
be written as the linear combination $\ket{q}=a_{0}\ket{0} + a_{1}\ket{1}$, where
$a_i \in \mathbb{C}$ and $\sum_i|a_i|^2=1$; this is a \emph{superposition} of the
two basis states, a concept with no analogy in classical computing.  Given the principal of 
superposition, an array of $n$ qubits can be in an equal superposition of all binary states 
from $\ket{0}$ unto $\ket{2^{n-1}}$, i.e. $\sum_{i=0}^{2^n-1}a_i\ket{\textsc{bin}(i)}$, 
where $a_i$ are complex numbers and $\textsc{bin}(i)$ is the binary expansion of
$i$. 

\subsubsection{Measurement:}

In quantum computing, \emph{measuring} a state is the only way to observe results
of calculation. Measuring an arbitrary quantum state $\ket{q}=a_0\ket{0} + a_1\ket{1}$
can result in two outcomes: $\ket{0}$ (with probability $|a_0|^2$), or $\ket{1}$
(with probability $|a_1|^2$). Moreover, the measurement will \emph{collapse} the
state leaving it in the state corresponding to the measurement result. 

The goal of a quantum algorithm is to manipulate the amplitudes of each 
binary state, $a_i$, such that the {\em incorrect} answers have very low amplitudes, 
$a_j \approx 0$, $ j = \text{incorrect}$ while the {\em correct} answers have 
amplitudes close to one, $a_j \approx 1$, $j = \text{correct}$.  This will ensure that 
after an algorithm is completed, we have a very high probability, when we measure 
every qubit, to measure the correct answer.   The simplest initial 
state is to initialise each qubit in the computer in the $\ket{+} = \left(\ket{0}+\ket{1}\right)/\sqrt{2}$ 
such that each $a_i = 1/2^{(n/2)}$, $\forall i$.  Therefore, initially, every possible 
binary state will have an equal probability of being measured.  The quantum algorithm will 
then manipulate these amplitudes to suppress the amplitudes of incorrect answers and 
increase the amplitude of correct ones.  At any given time the state of the quantum 
computer is represented by a $n$-dimensional complex vector $\ket{\psi} = (a_0,a_1,a_2,....,a_{2^{(n-1)}})^T$.

\subsubsection{Quantum gates:}

Quantum \emph{gates} act on qubits and modify their states and hence modify the 
amplitudes of each binary state, $a_i$. They are represented
as unitary (guaranteeing a gate is reversible, a necessity in quantum theory) 
matrices. An $n$-qubit
gate, $G$, is described by a $2^n \times 2^n$ matrix and its action on the state of the 
quantum computer is described by simply computing $\ket{\psi'} = G\ket{\psi}$, where 
$\ket{\psi'}$ is the output and $\ket{\psi}$ is the input.  It has been shown that any valid operation, 
$G$, can be decomposed into a discrete alphabet of single qubit and 2-qubit gates and 
consequently we only need to realise a small set of primitive qubit operations to realise 
any arbitrary computation.  Shown below is an example of such an alphabet, consisting of 
four single qubit gates and one two-qubit gate.  
\begin{equation}
\begin{aligned}
X &= 
\begin{pmatrix}
	0 & 1\\
	1 & 0
	\end{pmatrix}
Z= 
 \begin{pmatrix}
	1 & 0\\
	0 & -1
	\end{pmatrix}
H = 
\frac1{\sqrt2}\begin{pmatrix}
	1 & 1\\
	1 & -1
	\end{pmatrix}
\textsc{cnot}  =  
\begin{pmatrix}
	1 & 0 & 0 & 0\\
	0 & 1 & 0 & 0\\
	0 & 0 & 0 & 1\\
	0 & 0 & 1 & 0
	\end{pmatrix}
T  = 
\begin{pmatrix}
	1 & 0\\
	0 & e^{-i\pi/8}
	\end{pmatrix}
\end{aligned}
\end{equation}
These gates form a \emph{universal gate set} (technically, $\mathbb{S}=\{H,T,\textsc{cnot}\}$ are 
sufficient for universality, we include $X$ and $Z$ because of their relevance for QEC), i.e., arbitrary quantum gates can be
decomposed into products of these gates \cite{NC00}. (This is similar to the classical
case where all gates can be represented by equivalent circuits consisting of NAND
gates only.).

The properties of quantum information allow us to create certain states that have no classical analogue.  These 
states are called {\em entangled} states.  For example, if we prepare two qubits in the initial state $\ket{+}\ket{0}$ and 
apply the two qubit $\textsc{cnot}$ gate (where the control qubit is the one in the $\ket{+}$ state), we get the output 
$\ket{b}=\frac1{\sqrt2}(\ket{00}+\ket{11})$.  This state is known as a Bell state and it has properties that no classical 
computational state has.  Specifically if we measure one of the qubits in the $\ket{0}$ state, the second qubits is 
also found to be in the $\ket{0}$ state.  Similarly for the $\ket{1}$ state.  This behaviour is unique to quantum-bits and 
creation and manipulation of these types of states is an identifying feature when proving, experimentally, you have 
a true quantum system.  Entanglement is a fundamental property of quantum information and forms the 
basis of the TQC model we will discuss in the next section.  

\section{Topological Cluster State Computation}
\label{sec:topological}

\begin{figure*}[ht]
\begin{center}
\resizebox{60mm}{!}{\includegraphics{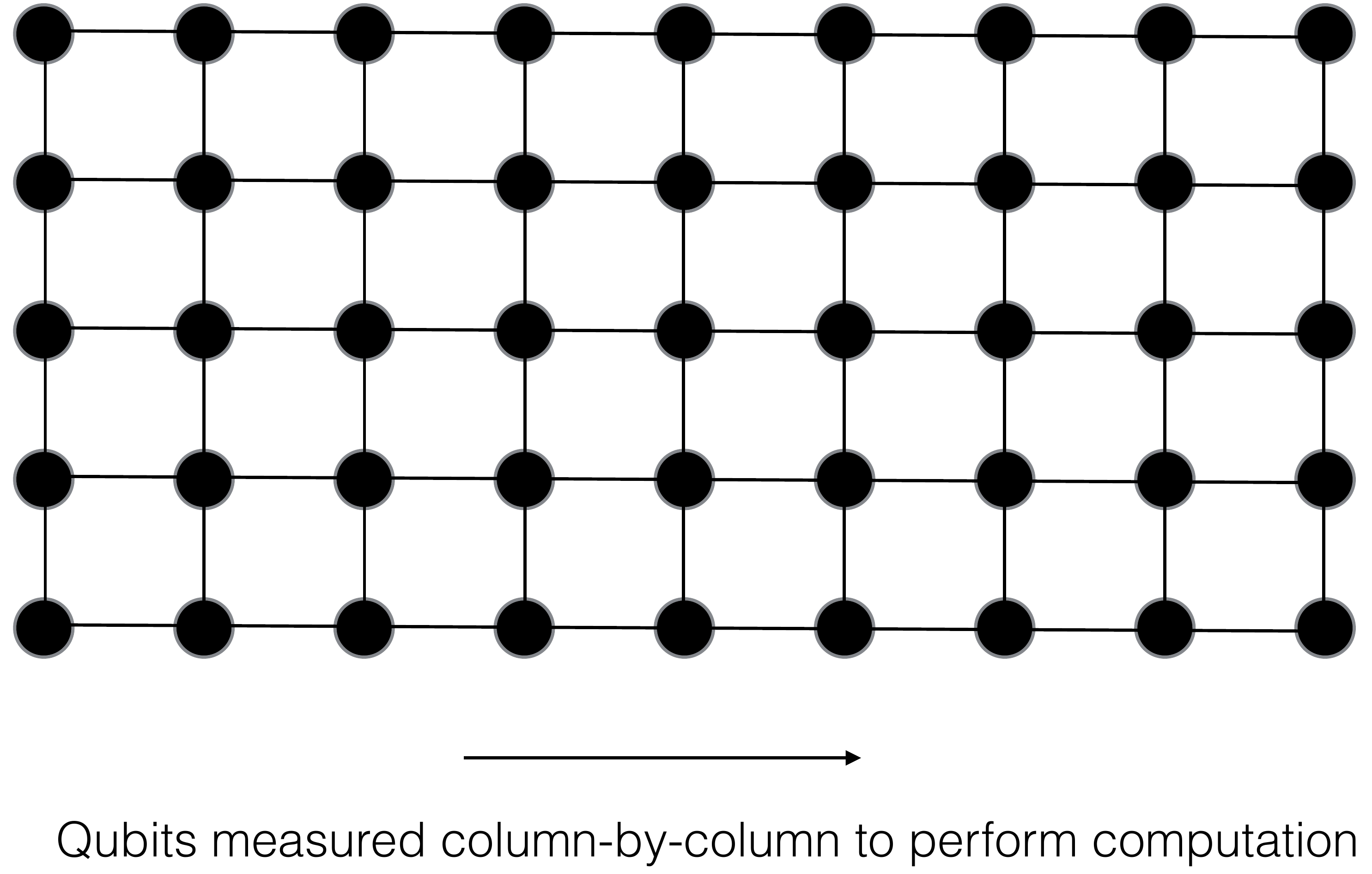}}
\end{center}
\vspace*{-10pt}
\caption{A standard 2D lattice of qubits used for measurement based quantum computation.  Qubits are measured from left to right 
and information is teleported from column to column.  Processing occurs during this teleportation, applying quantum gates.}
\label{fig:cluster}
\end{figure*}

The original formalism for quantum computation is the circuit based model \cite{NC00}.  This is where we have an array of 
qubits that is operated on by a pre-defined sequence of quantum gates to realise an algorithm.  There is another method of 
performing quantum computation, known as the measurement based model (MBM) \cite{RB01}.  In this model, we pre-define what is 
known as a Universal Resource State (URS).  A URS is a lattice of qubits where entanglement connections have been formed 
before any computation begins.  This URS can be thought of as a graph, where each vertex represents a qubit and each edge is 
a two-qubit quantum gate that establishes entanglement between two vertices.  Once this resource state has been prepared, 
quantum gates are realised by measuring individual qubits in well defined ways.  As computation proceeds, qubits are {\em 
consumed} as they are measured.  The first MBM was defined over a regular, 2-dimensional grid of qubits with nearest 
neighbour connections [Figure \ref{fig:cluster}].  In this model, qubits are measured, column-by-column, to realise quantum 
gates.  Essentially each {\em row} of qubits represented the world line of a given qubit of information and each {\em column} 
represented individual time steps of computation.  As each column is measured, information is teleported to the next column and 
a quantum gate is applied during this teleportation.  

This 2-dimensional MBM showed that arbitrary computation could be achieved using a pre-defined URS, but it did not 
incorporate any error correction protocols to protect against noise. 

\begin{figure*}[ht]
\begin{center}
\resizebox{70mm}{!}{\includegraphics{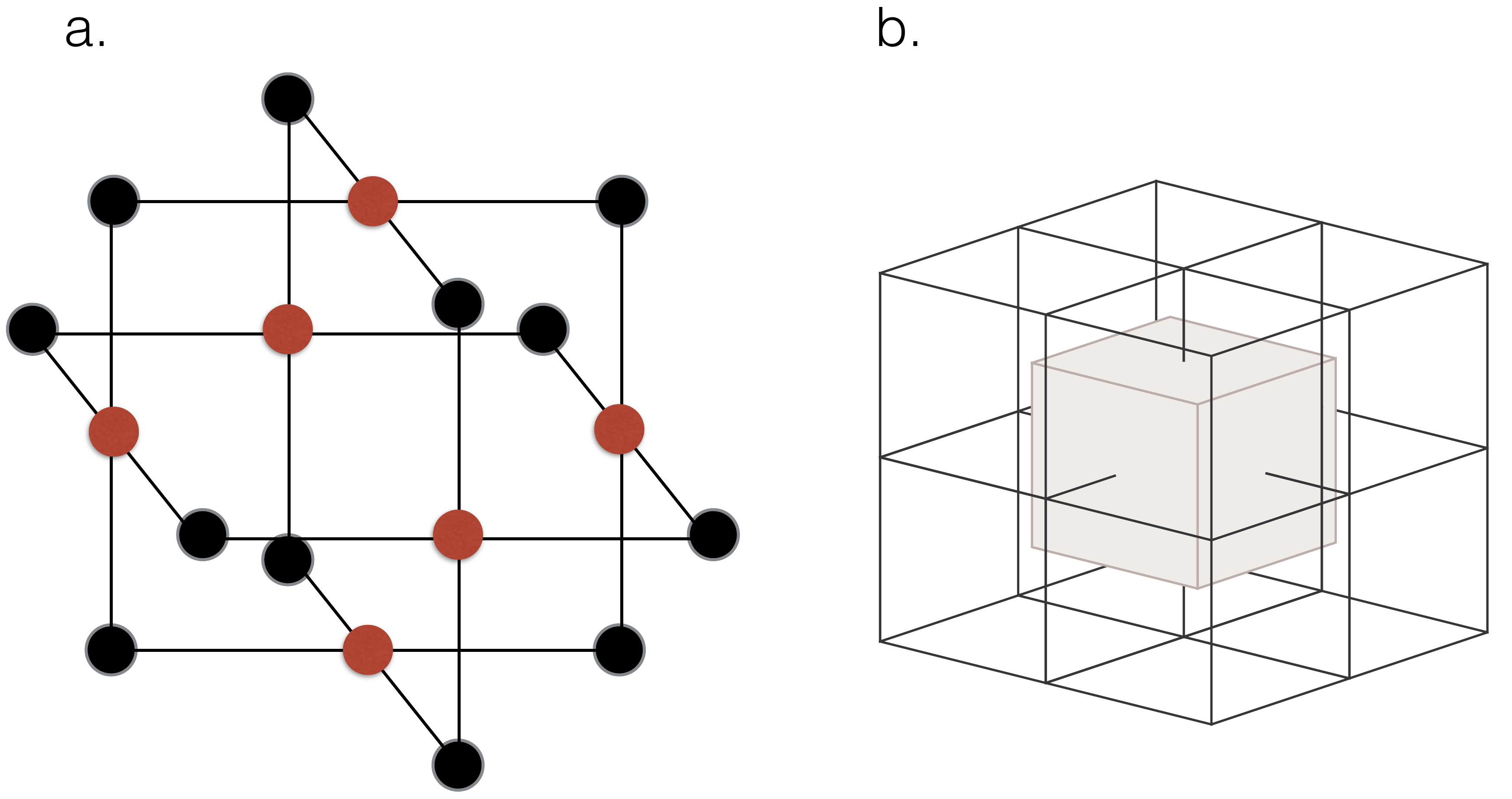}}
\end{center}
\vspace*{-10pt}
\caption{Figure a) represents the unit cell of the lattice.  Each of the Face qubits (red) are used to calculate the parity of the cell.  
The non-face qubits of Figure a) are face qubits on identical unit cells that 
are offset by half a lattice spacing along the three axes of the lattice.}
\label{fig:unit}
\end{figure*}

The Topological Cluster State model is a MBM of quantum computation that incorporates a sophisticated topological error 
correction model by construction.  It was derived from the seminal work of Kitaev \cite{K97} and extended to a 3-dimensional 
entangled lattice of qubits that forms the initial URS \cite{RHG07}. 
The fundamental unit cell of this lattice is illustrated in Figure \ref{fig:unit}.  Again, each vertex in the image represents a physical 
qubit while each edge represents a two-qubit gate applied to form an entanglement bond.  Preparing this state requires 
initialising each qubit in the $\ket{+}$ state, and applying a CZ gate between any two qubits connected by an edge.  A CZ gate 
can be achieved by applying the $\textsc{cnot}$ gate, interleaved by two $H$ gates on the target qubit \cite{NC00}.  
The total size of the 3-dimensional Topological cluster is dictated by the total resources needed for an algorithm.  i.e. how 
many encoded qubits and gates does the algorithm need and how strong the error correction needs to be to successfully 
complete computation.  For large quantum algorithms, the size of this lattice could be billions if not trillions of physical 
qubits \cite{DSMN13}.  

\subsection{Error Correction}
The primary job of the TQC model is to perform error correction.  The structure of the 3-dimensional 
lattice establishes certain symmetries that can be used to detect and correct errors that occur during the preparation and/or 
consumption of the state.  

Arbitrary noise on a qubit can be decomposed into a series of bit-flips ($X$ gates) and phase flips ($Z$ gates). A phase flip 
is a gate which can convert the state $\ket{+} = \left(\ket{0}+\ket{1}\right)/\sqrt{2}$ into $\ket{-}=\left(\ket{0}-\ket{1}\right)/\sqrt{2}$ and 
has no classical analogue.  A general error operator, $E$, acting on a single qubit can be written in the form,
\begin{equation}
E\ket{\psi} = k_I\ket{\psi} + k_xX\ket{\psi} + k_zZ\ket{\psi} + k_{xz}XZ\ket{\psi}
\end{equation}
where $\{|k_I|^2,|k_x|^2,|k_y|^2,|k_{xz}|^2\}$ are the probabilities that the qubit experiences an $X$ error, a $Z$ error or both \footnote{This is not a completely general description of a noise channel, but introducing the formalism for a general channel would require us to delve more into the mathematics of qubits.}.  Therefore, to protect qubits against noise, we just need the ability to 
detect and correct for bit- and phase-flips.  

\begin{figure*}[ht]
\begin{center}
\resizebox{100mm}{!}{\includegraphics{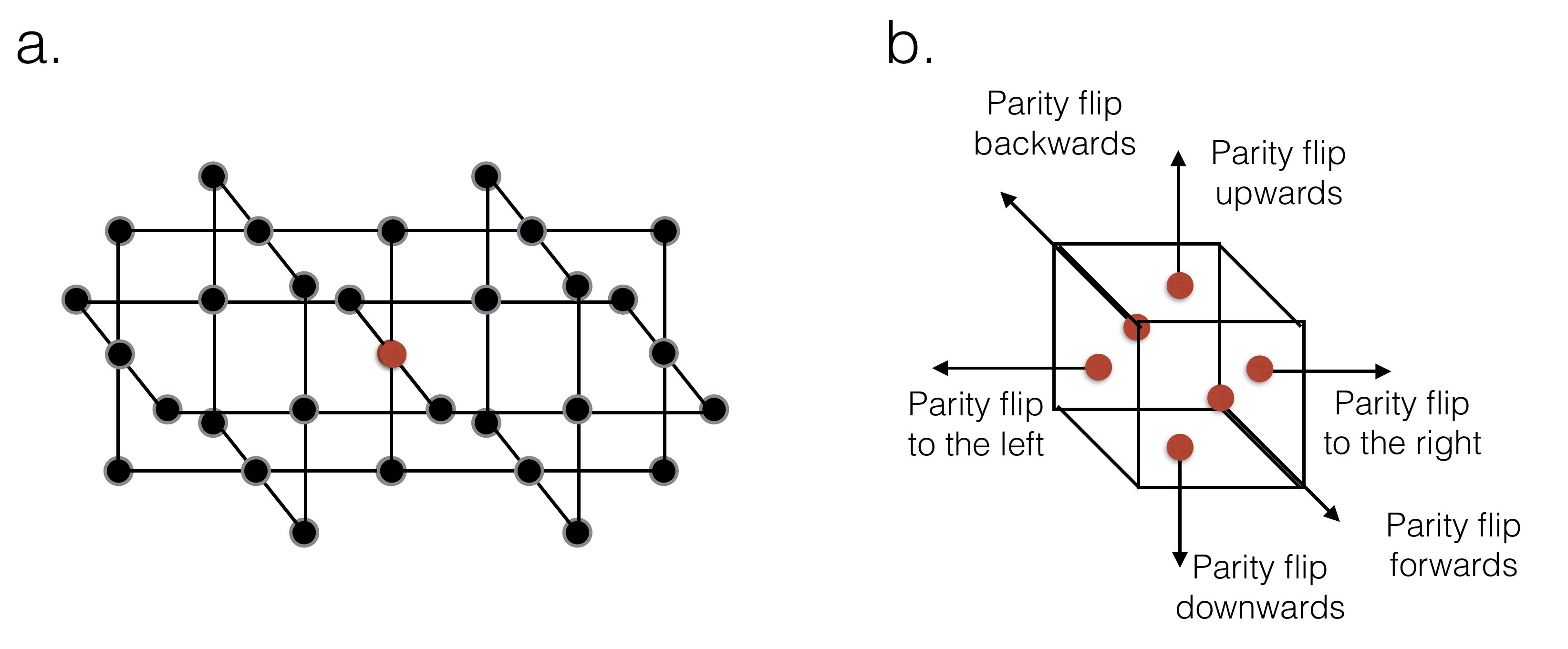}}
\end{center}
\vspace*{-10pt}
\caption{A single error on a face qubit of a unit cell will cause two parity flips on the cells which share the qubit [Figure a)].  The 
six neighbouring cells bordering a given cell allows us to uniquely determine which qubit experienced an error [Figure b)].}
\label{fig:error}
\end{figure*}

The unit cell of the topological cluster has certain symmetries.  Namely, if you measure the six face qubits of the unit cell 
(illustrated in red in Figure \ref{fig:unit}a)) in the basis $\{\ket{+},\ket{-}\}$ (known as an X-basis measurement) and you calculate 
the classical parity of the results (identifying the bit-value zero if we measure the qubit in $\ket{+}$ and one if we measure it in 
$\ket{-}$), you will always get an even parity result under modulo 2 addition.  
i.e. while the individual measurements themselves are random, the 
symmetries of the quantum state of the unit cell will conspire (through the property of entanglement) to always generate 
an even parity result when you combine the measured values of these six qubits.  Now, let us consider two of these unit cells 
side by side and the consequence of a $Z$-error on the qubit shared on a face [Figure \ref{fig:error}a)].  In quantum 
information the {\em order} in which you apply quantum gates is important.  For example, the output of the operation 
$XZ\ket{\psi}$ is not necessarily the same as the output of the operation $ZX\ket{\psi}$, this is because the gates $X$ and 
$Z$ do not {\em commute}, i.e. $XZ-ZX \neq 0$.  Instead, for these two operations the following holds, $XZ = -ZX$.  What 
does this mean when we measure our six face qubits of the unit cell when a qubit experiences an error?  If no error 
occurs, then the six measurement, when combined modulo 2, gives us an even parity result.  If one of those qubits 
experiences a $Z$-error prior to being measured in the $X$-basis the fact that $XZ = -ZX$ means that the measurement of 
the erred qubit will flip from $\ket{\pm}$ to $\ket{\mp}$.  Consequently, if the initial parity of the six measurements was  
even, it will flip to odd.  Hence for the two unit cells shown in Figure \ref{fig:error}a) when we measure the 11 face qubits 
and we observe a negative 
parity of the two sets of measurements, we can identify that a $Z$-error must have occurred on the qubit sharing a face 
between the two cells.  Similarly errors on the other five face qubits are detected by parity flips with the other unit cells 
bordering the five other faces [Figure \ref{fig:error}b)].  

An obvious question arises.  We have so far only considered the six qubits on each of the faces of the unit cell.  What about the 
other remaining qubits lying on edges?  If we 
stack together eight unit cells into a cube, at it's centre is an identical unit cell.  The face qubits associated with this unit cell 
correspond to the qubits on the edges of the eight cells in the cube.  The topological lattice embeds two self similar lattices, 
one which we call the primal lattice and the other which we call the dual\footnote{Which is primal and which is dual is arbitrary}.  
Face qubits on primal unit cells correspond to edge qubits on dual cells and visa versa.  These two self similar lattices also 
explains how $X$-errors are corrected.  In the previous paragraph we only considered $Z$ errors because the $Z$-gate didn't 
commute with the $X$-basis measurement of each face qubit and consequently the parity of the six face measurements flipped 
when an error occurred.  Again, without going into the mathematical detail, the symmetries of the topological lattice allows us 
to convert $X$-errors on a qubit into $Z$-errors on other qubits.  If an $X$-error occurs on a given qubit, the entanglement bonds 
connecting qubits can convert this $X$-error into $Z$-errors on all the qubits it is connected to \cite{FG08,G97+}.  If you examine the 
structure of the unit cell [Figure \ref{fig:unit}a)] you will note that a given face qubit is only connected to qubits on the edge of 
a unit cell.  Therefore an $X$-error occurring on a face qubit will be converted to $Z$-errors on edge qubits (which correspond 
to face qubits on dual cells).   Therefore, all errors can be converted to $Z$-errors in either the primal or dual lattices and 
detecting these parity flips in both spaces is sufficient for correcting arbitrary errors on each individual qubit.  

\begin{figure*}[ht]
\begin{center}
\resizebox{65mm}{!}{\includegraphics{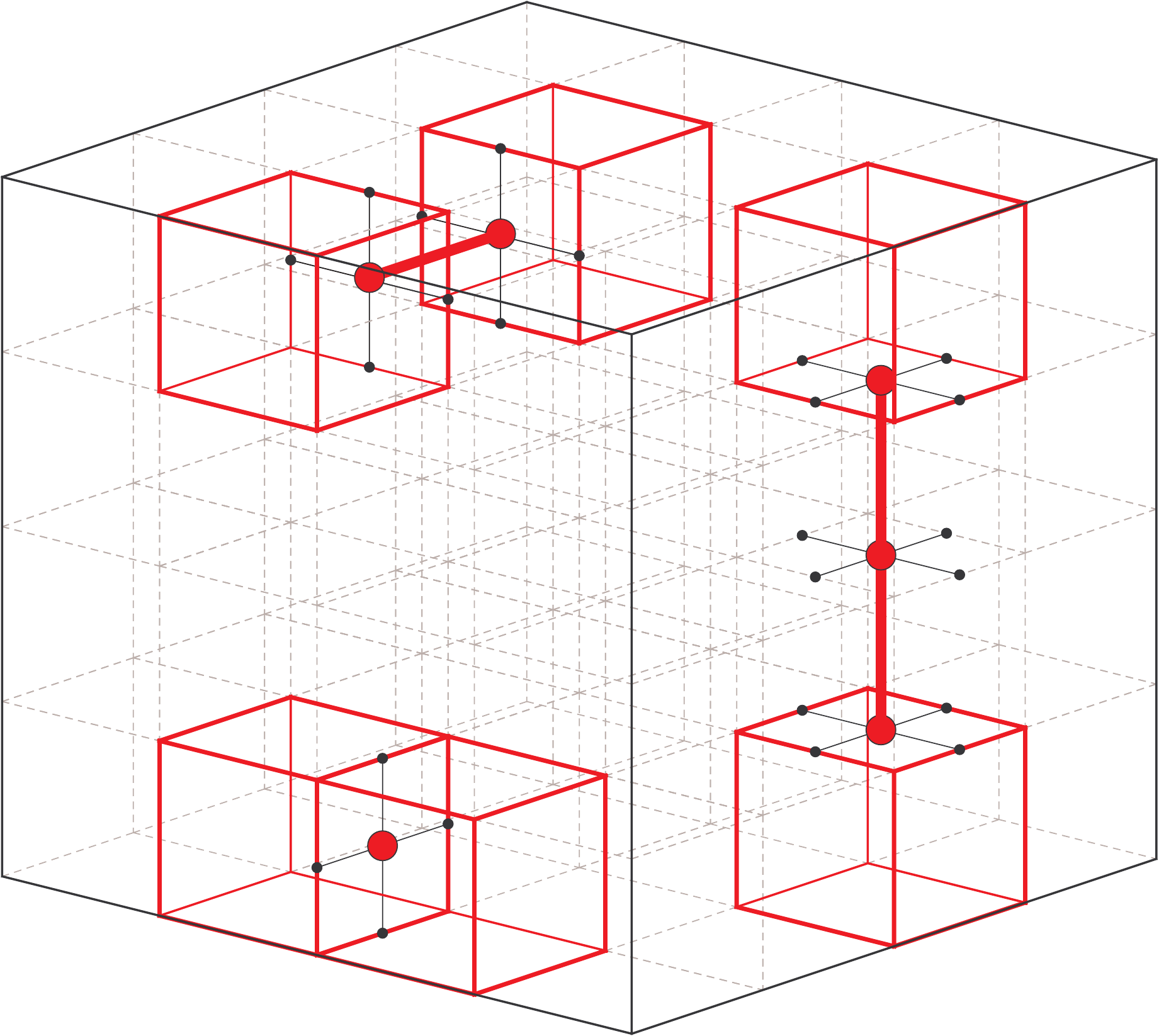}}
\end{center}
\vspace*{-10pt}
\caption{From Ref. \cite{FG08}.  Errors create parity flips on various unit cells.  Multiple errors can form chains.  Parity flips are 
only observed at the endpoints of chains.}
\label{fig:multiple}
\end{figure*}

We discussed how single errors can be corrected by examining the parity of neighbouring cells, the next issue is what happens 
when multiple errors occur.  This is shown in Figure \ref{fig:multiple}.  As the parity condition for a unit cell is calculated modulo two 
we only see an odd parity if an odd number of errors have occurred.  If an even number occur then the parity will remain even.  Therefore, 
if there is a chain of errors we will only see a parity flip for the two unit cells at the endpoint of the error chain.  In the case of isolated 
errors, endpoints are of neighbouring cells.  Hence decoding the error correction information requires us to match up the endpoints 
(which we detect via the calculation of a cells parity) with the actual physical sets of errors that occurred (which are not directly detected.  
 
In quantum information we assign a probability, $p$, that a given qubit will experience a bit ($X$) and/or phase ($Z$) error over some 
time interval, $t$.  This probability encapsulates the physical sources of noise such as environmental decoherence and control that 
could effect the operation of the qubit.  Provided that $p<1$, increasing numbers of errors occurring in a given time interval become 
exponentially less probable.  Consequently, the most probable event that gives rise to the observed set of parity flips in the topological 
cluster is the one with the fewest number of errors.  Given a set of parity flips measured in the topological cluster we connect them in 
a pairwise fashion such that the total length of all connections is minimised.  
This is a well known classical problem and was solved by Edmonds in 1967 \cite{E65} who developed a classical algorithm for minimum 
weight perfect matching who's runtime scales polynomially with the number of nodes (which in our case corresponds to the number 
of parity flips we observe).

\section{Physical Data Flow in an Operational Computer}  
What occurs in a physical quantum computer built using this model?  For the TQC model, the physical quantum hardware is 
responsible for preparing the lattice.  If we assume that the physical qubits in the quantum computer are single particles of light 
(photons), then each photon is prepared from a source and sent through the quantum computer to be entangled with its neighbours 
\cite{DFSG08}.  Each 2-dimensional cross-section of the lattice is prepared sequentially as photons "flow" through the quantum hardware.  

\begin{figure*}[ht]
\begin{center}
\resizebox{90mm}{!}{\includegraphics{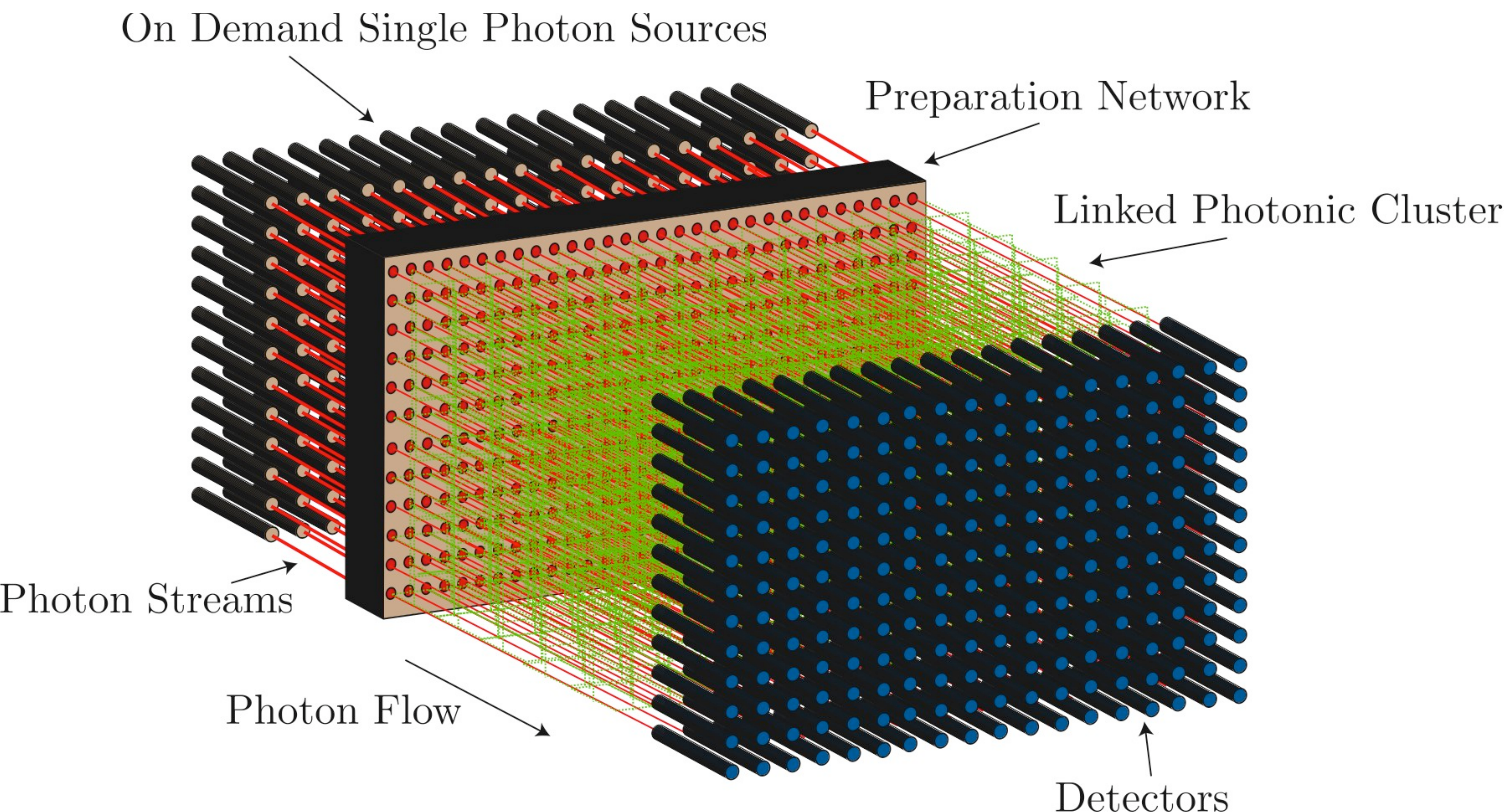}}
\end{center}
\vspace*{-10pt}
\caption{From Ref. \cite{DFTMN10}.  Architecture for an optical quantum computer.  Single photons are prepared, sent through a preparation network which 
is responsible for creating the topological lattice.  After the lattice is prepared it flows into detector arrays which performs 
measurement to perform computation.}
\label{fig:flow}
\end{figure*}

Photons are continuously injected into the rear of the preparation network.  
Each passes through a network of quantum devices, which act to 
link them together into the topological lattice.  Each quantum device operates on a fundamental 
clock cycle, $T$, and each device operates in a well-defined manner.  Once a given photon has been connected 
to its relevant neighbours, it does not have to wait until the rest of the lattice is constructed, it can be measured 
immediately.  This is exactly how the actual computer will operate.  The lattice is consumed at the same rate at which it 
is created, hence in the third dimension there only exists a small number of 2D cross-sections at any given time.

As one dimension of the topological lattice is identified as simulated time, the total 2D cross section 
defines the actual size of the quantum computer.  The lattice is built such that when each 2D cross-section is measured, all encoded
information is teleported 
to the next successive layer along the direction of simulated time allowing an algorithm to 
be implemented (in a similar manner to standard cluster state computation~\cite{RB01}).  

\begin{figure*}[ht]
\begin{center}
\resizebox{90mm}{!}{\includegraphics{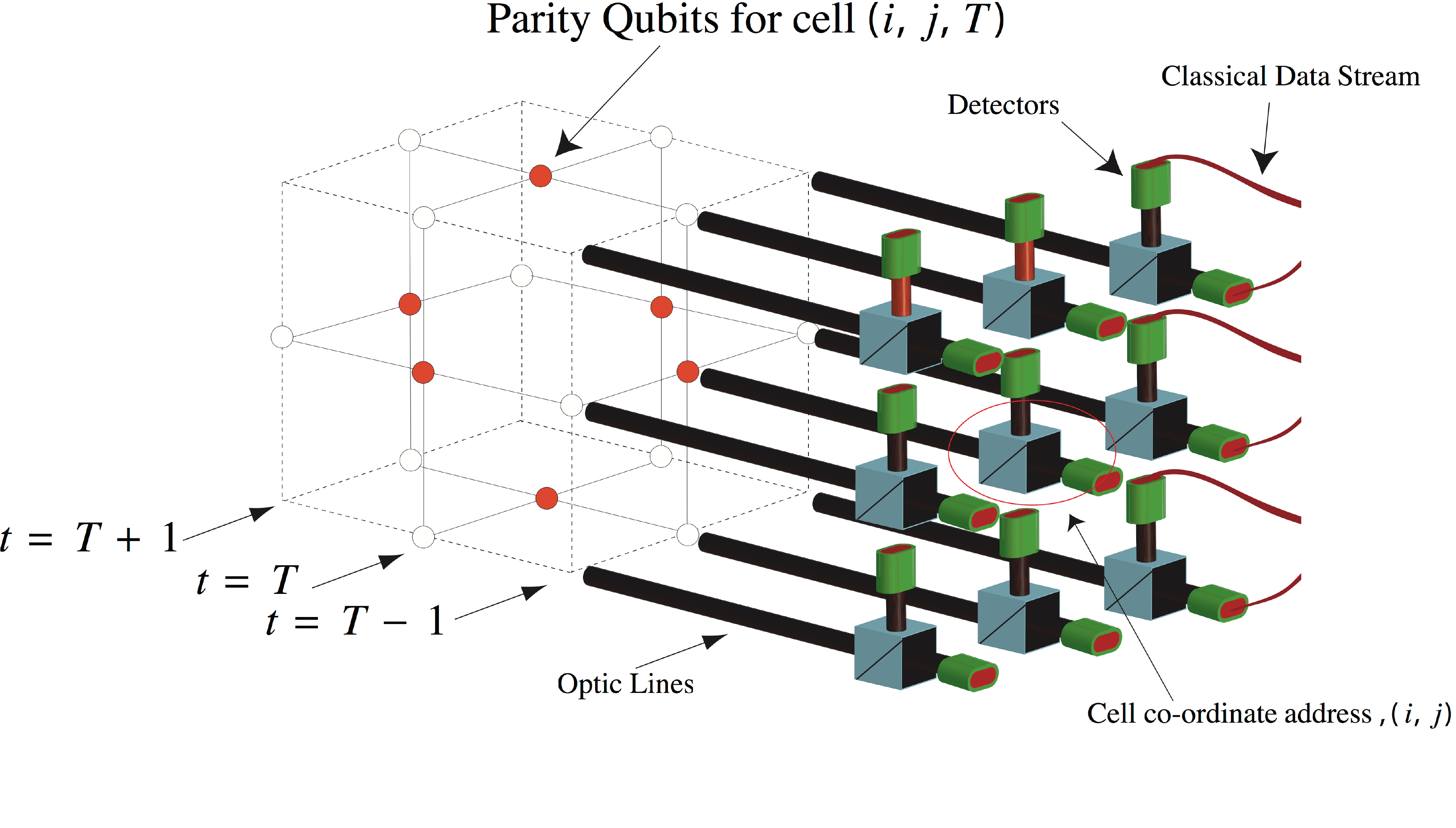}}
\end{center}
\vspace*{-10pt}
\caption{Detection array for a topological quantum computer where each qubit is a single photon.}
\label{fig:detection}
\end{figure*}

In Figure~\ref{fig:detection} we illustrate the structure of the detection system.  
A given unit cell flows through a set of nine optical fibres which carry the individual 
photons that have been linked together in the lattice.  As they flow into the detectors the parity of the 
cell is calculated as, 
\begin{equation}
P(i,j,T) = (s_{(i,j)}^{T-1} + s_{(i-1,j)}^T+s_{(i,j-1)}^T+s_{(i,j+1)}^T+s_{(i+1,j)}^T+s_{(i,j)}^{T+1}) \text { mod } 2 
\label{eq:parity}
\end{equation}
where $s_{i,j}^T$ is the detection result $(1,0)$ of detector $(i,j)$ at time $T$.  

Error decoding and correction must occur in real-time as the computer is operating in order to ensure 
the system will operate correctly.  Hence the classical data processing much be done efficiently, fast and 
in a highly parallel way.  

\section{The decoding problem}
\label{sec:errorcorrection}
The error correction decoding problem is a classical software and hardware optimisation problem to effectively perform the 
minimum weight perfect matching algorithm to an arbitrarily large topological lattice running at high speeds.  
Resource estimates for topological quantum computing has shown that to successfully implement fully error corrected, 
large-scale algorithms would require an enormous topological lattice \cite{DSMN13}.  The results of Ref. \cite{DSMN13} indicate that 
a lattice of the order of a billion cells in cross-section, running for a year at 10 nanoseconds per cross-sectional sheet 
is necessary to factor a 1024-bit number using Shor's algorithm. 
At 6-bits of raw data per cell, we would need to classically process on the order of $(6\times 10^9)/(30 \times 10^{-9}) = 2\times 10^{17}$ bits/second of data to perform error correction decoding for the entire computation.  

\begin{figure*}[ht]
\begin{center}
\resizebox{80mm}{!}{\includegraphics{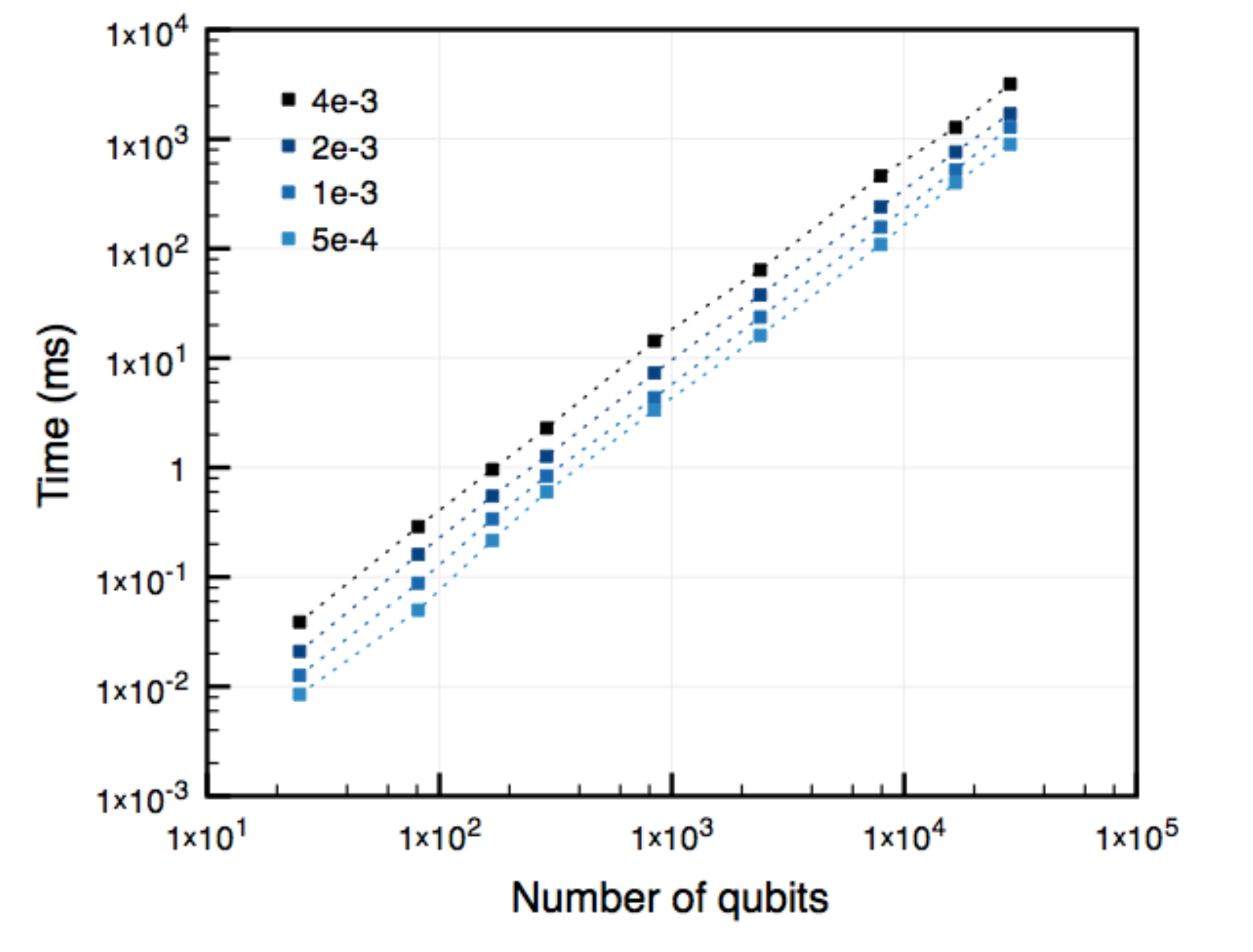}}
\end{center}
\vspace*{-10pt}
\caption{Processing speed of Blossom V \cite{K08} as a function of the total number of qubits in a 2D cross-section of the 
topological lattice.  Each curve represents a different physical error rate of each qubit, $p$.  This plot was produced with 
a standard laptop with no further optimisation.}
\label{fig:simulate}
\end{figure*}

\begin{figure}[ht!]
\begin{center}
\resizebox{70mm}{!}{\includegraphics{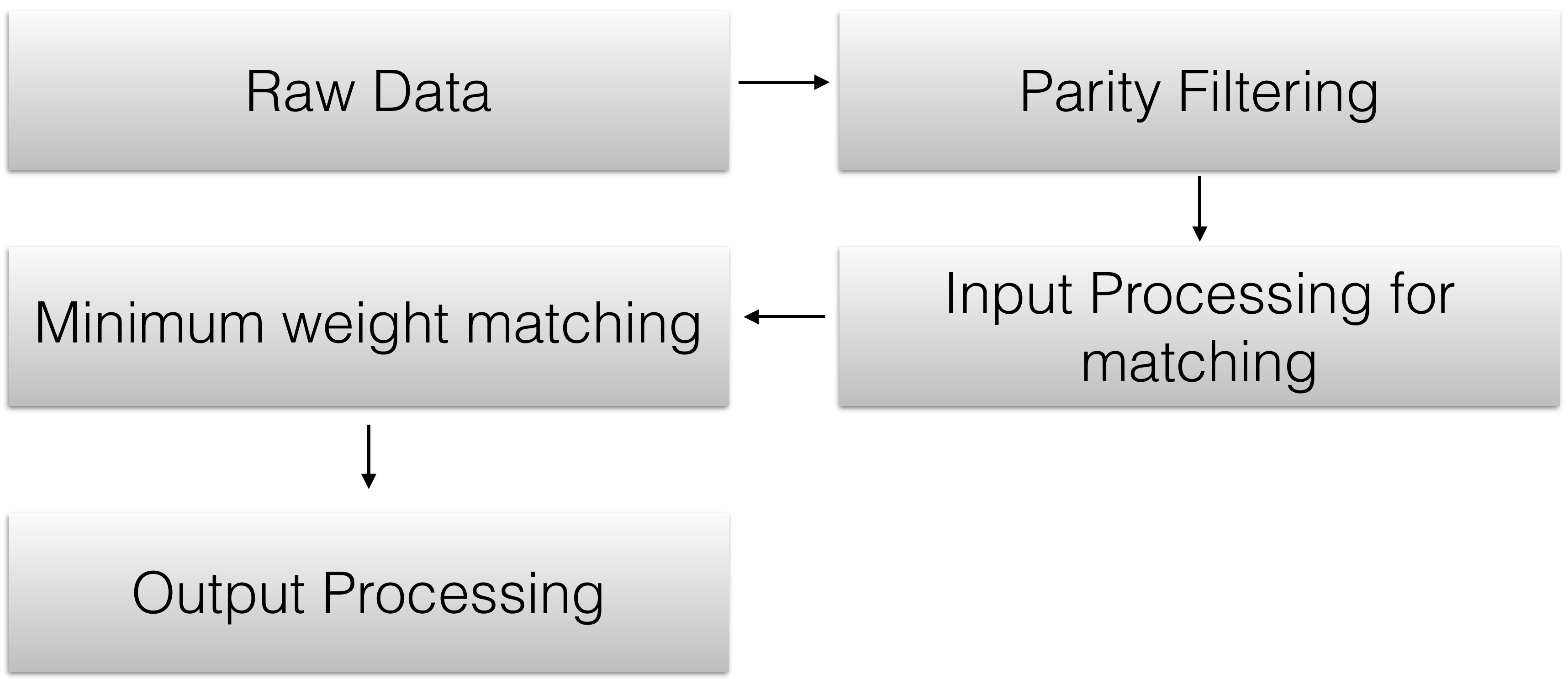}}
\end{center}
\vspace*{-10pt}
\caption{Processing stages for error decoding in the topological model.}
\label{fig:processing}
\end{figure}

This clearly is a phenomenal amount of data that needs to be processed while the computer is running.  Clearly we require a large 
amount of parallel processing and a modular classical processing framework to decode error correction data for a full-scale 
machine.   There has been work attempting to address this problem  which falls into two categories.  The first is further 
optimisation of the minimum weight perfect matching algorithm.  The Blossom V algorithm is currently used when performing 
simulations of the topological cluster state model \cite{K08} and we can examine its performance for large lattices [Figure \ref{fig:simulate}].  
From this figure (which was produced by running the algorithm on a standard laptop) shows that Blossom V runs far too slowly to handle 
the processing of error correction data for a large-scale computer.  This necessitates further optimisation of the algorithm.  Work by 
Fowler and others \cite{FWH12,F13} attempts to rectify this problem, but at this stage no benchmarking has been performed using this package.  The 
second category is dedicated hardware implementations of the decoding operations \cite{DFTMN10}.  There are several steps which is illustrated in 
Figure. \ref{fig:processing}.  

The raw data is the bit streams coming directly from the quantum hardware.  Parity filtering is 
the first step, where the co-ordinates of unit cells that have experienced a parity flip are retained and all other data is disregarded.  
This can reduce the amount of information as the probability that a unit cell of the lattice will experience a parity flip is of the order of 
the error rate of each qubit, $p$, which will be approximately 0.1\% \cite{S14}.  The next step is to convert the collection of co-ordinates 
into a graph which is used as input for the minimum weight matching algorithm.  This data will produce a lookup table associating a 
vertex number for the graph with the co-ordinate of the relevant cell.  The matching algorithm comes next and will produce a list of bi-partite connections telling us which nodes in the graph are connected.  Output processing then converts these nodes back into the 
cell co-ordinates allowing us to correct the actual errors.  

Each of these stages will have to be handled by dedicated circuits, built primarily for speed.  This has not currently been done and 
we do not have evidence if current technology is sufficient to achieve fast enough speeds for 
quantum computing systems.  For various physics related reasons, we do not wish to slow down the operational speed of the 
quantum hardware to accommodate slow classical processing.  The speed of the classical system mush be commensurate with the 
quantum system (which can vary between 10ns and 10ms depending on the underlying technology).  The first generation of quantum 
computers will be slow, so the demands on the classical hardware should not be too significant in the short term.  But more futuristic 
technology is being developed \cite{K98} and will run at much higher clock rates.  Designing the classical system with these faster 
systems in mind should ensure that quantum computer development is not bottlenecked by the necessary classical systems being 
underdeveloped.  

\section{Acknowledgements}
We wish to thank Ashley 
Stephens for producing the image in Figure \ref{fig:simulate}.  

\small
\bibliographystyle{splncs}
\bibliography{bib1.bib}

\begin{thebibliography}{10}

\bibitem{NC00}
Nielsen, M., Chuang, I.:
\newblock {Quantum Computation and Information}. Second edn.
\newblock Cambridge University Press (2000)

\bibitem{G06}
Gaebel, T., Domhan, M., Popa, I., Wittmann, C., Neumann, P., Jelezko, F.,
  Rabeau, J., Stavrias, N., Greentree, A., Prawer, S., Meijer, J., Twamley, J.,
  Hemmer, P., Wrachtrup, J.:
\newblock {Room Temperature coherent control of coupled single spins in solid}.
\newblock Nature Physics (London) \textbf{2} (2006)  408--413

\bibitem{HA08}
Hanson, R., Awschalom, D.:
\newblock {Coherent manipulation of single spins in semiconductors}.
\newblock Nature (London) \textbf{453} (2008)  1043--1049

\bibitem{PLZY08}
Press, D., Ladd, T.D., Zhang, B., Yamamoto, Y.:
\newblock {Complete quantum control of a single quantum dot spin using
  ultrafast optical pulses}.
\newblock Nature (London) \textbf{456} (2008)  218--221

\bibitem{PMO09}
Politi, A., Matthews, J., O'Brien, J.:
\newblock {Shor's quantum factoring algorithm on a photonic chip}.
\newblock Science \textbf{325} (2009)  1221

\bibitem{P12}
Pla, J., Tan, K.Y., Dehollain, J.P., Lim, W.H., Morton, J.J.L., Jamieson, D.N.,
  Dzurak, A.S., Morello, A.:
\newblock {A single-atom electron spin qubit in Silicon}.
\newblock Nature (London) \textbf{489} (2012)  541--545

\bibitem{L12}
Lucero, E., Barends, R., Chen, Y., Kelly, J., Mariantoni, M., Megrant, A.,
  O'Malley, P., Sank, D., Vainsencher, A., Wenner, J., White, T., Yin, Y.,
  Cleland, A.N., Martinis, J.:
\newblock {Computing prime factors with a Josephson phase qubit quantum
  processor}.
\newblock Nature Physics \textbf{8} (2012)  719--723

\bibitem{L10}
Ladd, T., Jelezko, F., Laflamme, R., Nakamura, Y., Monroe, C., O'Brien, J.:
\newblock {Quantum Computing}.
\newblock Nature (London) \textbf{464} (2010)  45--53

\bibitem{VAMWL14}
Vinci, W., Albash, T., Mishra, A., Warburton, P.A., Lidar, D.A.:
\newblock {Distinguishing Classical and Quantum Models for the D-Wave Device}.
\newblock arxiv:1403.4228 (2014)

\bibitem{BRIWWLMM13}
Boixo, S., Ronnow, T.F., Wecker, S.I.Z.W.D., Lidar, D., Martinis, J., Troyer,
  M.:
\newblock {Quantum annealing with more than one hundred qubits}.
\newblock Nature Physics \textbf{10} (2014)  218

\bibitem{SSSV14}
Shin, S., Smith, G., Smolin, J., Vazirani, U.:
\newblock {How "Quantum" is the D-Wave Machine?}
\newblock arxiv:1401.0787 (2014)

\bibitem{B14}
Barends, R., Kelly, J., Megrant, A., Veitia, A., Sank, D., Jeffrey, E., White,
  T., Mutus, J., Fowler, A., Campbell, B., Chen, Y., Chen, Z., Chiaro, B.,
  Dunsworth, A., Neill, C., O`Malley, P., Roushan, P., Vainsencher, A., Wenner,
  J., Korotkov, A., Cleland, A., Martinis, J.:
\newblock {Logic gates at the surface code threshold: Superconducting qubits
  poised for fault-tolerant quantum computing}.
\newblock arXiv:1402.4848 (2014)

\bibitem{CDMF14}
Choi, T., Debnath, S., Manning, T., Figgatt, C., Gong, Z.X., Duan, L.M.,
  Monroe, C.:
\newblock {Optimal quantum control of multi-mode couplings between trapped ion
  qubits for scalable entanglement}.
\newblock arxiv:1401.1575 (2014)

\bibitem{DMN13}
Devitt, S., Munro, W., Nemoto, K.:
\newblock Quantum error correction for beginners.
\newblock Rep. Prog. Phys. \textbf{76} (2013)  076001

\bibitem{S03+}
Steane, A.:
\newblock {Quantum Computing and Error Correction}.
\newblock {Decoherence and its implications in quantum computation and
  information transfer, Gonis and Turchi, eds, pp.284-298 (IOS Press,
  Amsterdam, 2001), quant-ph/0304016} (2001)

\bibitem{CRSS98}
Calderbank, A., Rains, E., Shor, P., Sloane, N.:
\newblock {Quantum Error Correction via Codes Over GF(4)}.
\newblock IEEE Trans. Inform. Theory \textbf{44} (1998)  1369

\bibitem{KLV99}
Knill, E., Laflamme, R., Viola, L.:
\newblock {Theory of Quantum Error Correction for General Noise}.
\newblock Phys. Rev. Lett. \textbf{84} (2000)  2525

\bibitem{AB97}
Aharonov, D., Ben-Or, M.:
\newblock {Fault-tolerant Quantum Computation with constant error}.
\newblock Proceedings of 29th Annual ACM Symposium on Theory of Computing
  (1997) ~46

\bibitem{K97}
Kitaev, A.:
\newblock {Quantum Computations: algorithms and error correction}.
\newblock Russ. Math. Serv. \textbf{52} (1997)  1191

\bibitem{DKLP02}
Dennis, E., Kitaev, A., Landahl, A., Preskill, J.:
\newblock {Topological Quantum Memory}.
\newblock J. Math. Phys. \textbf{43} (2002)  4452

\bibitem{RHG06}
Raussendorf, R., Harrington, J., Goyal, K.:
\newblock {A Fault-tolerant one way quantum computer}.
\newblock Ann. Phys. \textbf{321} (2006)  2242

\bibitem{RH07}
Raussendorf, R., Harrington, J.:
\newblock {Fault-tolerant quantum computation with high threshold in two
  dimensions}.
\newblock Phys. Rev. Lett. \textbf{98} (2007)  190504

\bibitem{FMMC12}
Fowler, A., Mariantoni, M., Martinis, J., Cleland, A.:
\newblock {Surface codes: Towards practical large-scale quantum computation}.
\newblock Phys. Rev. A. \textbf{86} (2012)  032324

\bibitem{DFTMN10}
Devitt, S., Fowler, A., Tilma, T., Munro, W., Nemoto, K.:
\newblock {Classical Processing Requirements for a Topological Quantum
  Computing Systems}.
\newblock Int. J. Quant. Inf. \textbf{8} (2010) ~1

\bibitem{GP14}
Duclos-Cianci, G., Poulin, D.:
\newblock {Fault-Tolerant Renormalization Group Decoded for Abelian Topological
  Codes}.
\newblock Quant. Inf. Comp. \textbf{14} (2014)  0721

\bibitem{KMM13}
Kliuchnikov, V., Maslov, D., Mosca, M.:
\newblock {Asymptotically optimal approximation of single qubit unitaries by
  Clifford and T circuits using a constant number of ancillary qubits}.
\newblock Phys. Rev. Lett. \textbf{110} (2013)  190502

\bibitem{VI05}
Meter, R.V., Itoh, K.:
\newblock {Fast Quatum Modular Exponentiation}.
\newblock Phys. Rev. A. \textbf{71} (2005)  052320

\bibitem{Z98}
Zalka, C.:
\newblock {Fast Versions of Shor's quantum factoring algorithm}.
\newblock quant-ph/9806084 (1998)

\bibitem{VBE96}
Vedral, V., Barenco, A., Ekert, A.:
\newblock {Quantum Networks for elementary arithmetic operations}.
\newblock Phys. Rev. A. \textbf{54} (1996)  147

\bibitem{CV11}
Choi, B., Meter, R.V.:
\newblock {A $\Theta(\sqrt{n})$-depth Quantum Adder on a 2D NTC Quantum
  Computer Architecture}.
\newblock ACM Journal on Emerging Technologies in Computer Systems (JETC)
  \textbf{7} (2011) ~11

\bibitem{WC00}
Cleve, R., Watrous, J.:
\newblock {Fast Parallel circuits for the quantum fourier transform}.
\newblock {Proc. 41st Annual IEEE Symposium on Foundations of Computer Science
  (FOCS 2000)} (2000)  526--536

\bibitem{MI05}
Meter, R.V., Itoh, K.:
\newblock {Fast Quantum Modular Exponentiation}.
\newblock Phys. Rev. A. \textbf{71} (2005)  052320

\bibitem{RHG07}
Raussendorf, R., Harrington, J., Goyal, K.:
\newblock {Topological fault-tolerance in cluster state quantum computation}.
\newblock New J. Phys. \textbf{9} (2007)  199

\bibitem{FG08}
Fowler, A., Goyal, K.:
\newblock {Topological cluster state quantum computing}.
\newblock Quant. Inf. Comp. \textbf{9} (2009)  721

\bibitem{SJ08}
Stock, R., James, D.:
\newblock {A Scalable, high-speed measurement based quantum computer using
  trapped ions}.
\newblock Phys. Rev. Lett. \textbf{102} (2009)  170501

\bibitem{DFSG08}
Devitt, S., Fowler, A., Stephens, A., Greentree, A., Hollenberg, L., Munro, W.,
  Nemoto, K.:
\newblock {Architectural design for a topological cluster state quantum
  computer}.
\newblock New. J. Phys. \textbf{11} (2009)  083032

\bibitem{NTDS13}
Nemoto, K., Trupke, M., Devitt, S., Stephens, A., Buczak, K., Nobauer, T.,
  Everitt, M., Schmiedmayer, J., Munro, W.:
\newblock {Photonic architecture for scalable quantum information processing in
  NV-diamond}.
\newblock arXiv:1309.4277 (2013)

\bibitem{JMFMKLY10}
Jones, N.C., Meter, R.V., Fowler, A., McMahon, P., Kim, J., Ladd, T., Yamamoto,
  Y.:
\newblock {A Layered Architecture for Quantum Computing Using Quantum Dots}.
\newblock Phys. Rev. X. \textbf{2} (2012)  031007

\bibitem{MF10}
Meter, R.V., Ladd, T., Fowler, A., Yamamoto, Y.:
\newblock {Distributed Quantum Computation Architecture Using Semiconductor
  Nanophotonics}.
\newblock Int. J. Quant. Inf. \textbf{8} (2010)  295

\bibitem{MRRBMD14}
Monroe, C., Raussendorf, R., Ruthven, A., Brown, K., Maunz, P., Duan, L.M.,
  Kim, J.:
\newblock {Large Scale Modular Quantum Computer Architecture with Atomic Memory
  and Photonic Interconnects}.
\newblock Phys. Rev. A. \textbf{89} (2014)  022317

\bibitem{RB01}
Raussendorf, R., Briegel, H.J.:
\newblock {A One way Quantum Computer}.
\newblock Phys. Rev. Lett. \textbf{86} (2001)  5188

\bibitem{DSMN13}
Devitt, S., Stephens, A., Munro, W., Nemoto, K.:
\newblock Requirements for fault-tolerant factoring on an atom-optics quantum
  computer.
\newblock Nature Communications \textbf{4} (2013)  2524

\bibitem{G97+}
Gottesman, D.:
\newblock {PhD Thesis (Caltech)}.
\newblock quant-ph/9705052 (1997)

\bibitem{E65}
Edmonds, J.:
\newblock {Paths, trees, and flowers}.
\newblock Canadian J. Math. \textbf{17} (1965)  449

\bibitem{K08}
Kolmogorov, V.:
\newblock {Blossom V: A new implementation of a minimum cost perfect matching
  algorithm}.
\newblock Math. Prog. Comp. \textbf{1} (2009) ~43

\bibitem{FWH12}
Fowler, A., Whiteside, A., Hollenberg, L.:
\newblock {Towwards practical classical processing for the surface code: Timing
  analysis}.
\newblock Phys. Rev. A. \textbf{86} (2012)  042313

\bibitem{F13}
Fowler, A.:
\newblock {Minimum weight perfect matching in O(1) parallel time}.
\newblock arxiv:1307.1740 (2013)

\bibitem{S14}
Stephens, A.:
\newblock {Fault-tolerant thresholds for quantum error correction with the
  surface code}.
\newblock Phys. Rev. A. \textbf{89} (2014)  022321

\bibitem{K98}
Kane, B.:
\newblock {A Silicon-Based nuclear spin Quantum Computer}.
\newblock Nature (London) \textbf{393} (1998)  133

\end{thebibliography}

\end{document}